\documentclass[%
 reprint,
 amsmath,amssymb,
 aps,
]{revtex4-2}

\usepackage{graphicx}
\usepackage{dcolumn}
\usepackage{bm}

\usepackage{float}

\usepackage{enumitem}

\usepackage{fancyhdr}

\begin{document}

\preprint{APS/123-QED}

\title{Evidences of conformal invariance in 2d rigidity percolation}

\author{Nina Javerzat}
 \email{njaverza@sissa.it}
\affiliation{
 SISSA and INFN Sezione di Trieste, via Bonomea 265, 34136, Trieste, Italy
}%

\author{Mehdi Bouzid}
 \email{mehdi.bouzid@univ-grenoble-alpes.fr}
\affiliation{%
Univ. Grenoble Alpes, CNRS, Grenoble INP, 3SR, F-38000, Grenoble, France
}%

\date{\today}

\begin{abstract}
The rigidity transition occurs when, as the density of microscopic components is increased, a disordered medium becomes able to transmit and ensure macroscopic mechanical stability, owing to the appearance of a space-spanning rigid connected component, or cluster. As a continuous phase transition it exhibits a scale invariant critical point, at which the rigid clusters are random fractals.
We show, using numerical analysis, that these clusters are also conformally invariant, and we use conformal field theory to predict the form of universal finite size effects. Furthermore, although connectivity and rigidity percolation are usually though to belong to different universality classes and thus be of fundamentally different natures, we provide evidence of unexpected similarities between the statistical properties of their random clusters at criticality. 
Our work opens a new research avenue through the application of the powerful 2D conformal field theory tools to understand the critical behavior of a wide range of physical and biological materials exhibiting such a mechanical transition.

\end{abstract}

\maketitle

\textit{Introduction} -- 
Symmetries are the cornerstone to understand and to model physical phenomena~\cite{gross1996role}, and their identification, a powerful guiding principle for deriving physical laws. Indeed, the compatibility between symmetries often results in constraints on the physical properties of the system: for example the compatibility of discrete translations and rotations in crystals leads to the crystallographic restriction theorem, which classifies all patterns of periodic discrete lattices one can encounter in nature~\cite{giacovazzo2011fundamentals}. But symmetries are not only deterministic: 
second order phase transitions are a paradigmatic example of systems possessing a symmetry of random nature,  where the long-range statistical fluctuations are invariant in law under change of scale. For a host of systems exhibiting critical behaviour --as diverse as linear polymers \cite{Saleur_1987}, graphene membranes \cite{giordanelli2016conformal}, disordered systems \cite{Delfino2021-fv},  a larger symmetry emerges and fluctuations are also invariant under local rescalings i.e. under all geometrical transformations that preserve angles and rescale distances, called conformal transformations \cite{Polyakov1970}.
The emergence of this enhanced symmetry is a powerful tool: exploiting the compatibility constraints on the physical observables allows to understand and predict the universal features of phase transitions \cite{Poland2016}, and even in some cases to fully characterise the scaling limit \cite{BPZ}. 
The origin of conformal symmetry is however still not systematically understood \cite{Nakayama2013}, even in two dimensions. Indeed, while in 2d unitary systems conformal invariance is automatically implied by scale invariance \cite{polchinski88,Zamo86} this is not anymore true for non-unitary phenomena, of which percolation is maybe the most representative and versatile example. Still, percolation in its various forms is believed (in some cases proven) to be conformally invariant, for instance: uncorrelated (Bernoulli) percolation \cite{smirnov2001}, the random $Q-$states Potts model \cite{smirnov-lattmodels}, percolation of random surfaces \cite{giordanelli2016conformal,javerzat2020topological}, and to our knowledge there is no equilibrium percolation model which has been shown to be scale but not conformal invariant.\\
In this context, rigidity percolation (RP) is an ideal model to study the possible emergence of conformal symmetry. On the one hand, establishing the conformal invariance of this phase transition, of prominent importance in soft matter, may allow to better characterise its still poorly known universality class. On the other hand, it is the first time that conformal invariance is studied in a percolation phenomenon of mechanical nature (a priori distinct from the "connectivity percolation" (CP) models mentioned above), and this might shed some light on which features of a percolation model make its scaling limit conformally invariant.

Rigidity percolation in central force random springs models provides a generic theoretical and simple framework to study how a system transitions from a liquid to a solid phase, where the underlying building blocks assemble into a percolating cluster that is able to transmit stresses to the boundary and sustain external loads. It has been successfully used to highlight the structural and mechanical properties of many soft materials such as living tissues~\cite{petridou2021rigidity}, biopolymers networks~\cite{broedersz2011criticality,mao2015mechanical}, molecular glasses ~\cite{thorpe2000self}, stability of granular packings~\cite{feng1985percolation,henkes2016rigid,berthier2019rigidity} or colloidal gelation~\cite{mehdi19,rouwhorst2020nonequilibrium,tsurusawa2019direct}.
 Several critical exponents, characterising the long-distance critical behaviour, have been numerically determined, such as the correlation length exponent $\nu = 1.21\pm 0.06$ and the order parameter exponent $\beta = 0.18 \pm 0.02$, defining an a priori new universality class \cite{jacobs1995generic}. Hyperscaling relations also give the fractal dimension of the rigid cluster as $d_f = 2-\beta/\nu = 1.86\pm0.02 $, a value which was confirmed by direct measurement ~\cite{mehdi19}.

In this article, we show that the rigidity percolation clusters exhibit conformal invariance at the critical point and, interestingly, that the fine statistical properties of the RP clusters and of the CP clusters share surprising similarities, despite belonging to distinct universality classes.

\textit{Model and Methods -- }
We perform three independent numerical tests of conformal invariance, based on the study of a geometrical property of the random rigid clusters, their so-called $n-$point connectivity \cite{Grimmett_1999}:
\begin{equation}\label{eq:defconn}
p_{12\cdots n}(z_1,\cdots,z_n)\overset{\mathrm{def}}{=}\mathrm{Prob}\left[z_1,\cdots,z_n\in\;\mathcal{RC}\right].
\end{equation}
$z_i$ are points in the two-dimensional space and $\mathcal{RC}$ denotes a rigid cluster. (\ref{eq:defconn}) gives therefore the probability that $n$ points are connected by paths inside the same rigid cluster. These quantities have been very useful to understand connectivity percolation \cite{Delfino2010,Picco2016,Jacobsen2018,He2020, Javerzat2020}. We make the central assumption that, in the scaling limit, the connectivities (\ref{eq:defconn}) can be described by a field theory, and more precisely that they are given by correlation functions of a scaling field that we denote $\Phi_c$, of scaling dimension $\Delta_c$:
\begin{equation}\label{eq:assumption}
    p_{12\cdots n}(z_1,\cdots,z_n) \;
    \overset{\scriptscriptstyle \mathrm{scaling}}{ \underset{\scriptscriptstyle\mathrm{lim.}}{\rightarrow}}\; a_0^{(n)}\,\langle \Phi_c(z_1)\cdots \Phi_c(z_n)\rangle
\end{equation} 
where $a_0^{(n)}$ is a non-universal constant that depends on the microscopic details of the model.

When present, conformal symmetry constrains the form of correlations, hence of the connectivities, in a precise way.
In this work we use a lattice model of rigidity percolation to measure numerically certain rigid cluster connectivities on specific geometries. Using (\ref{eq:assumption}) gives the corresponding CFT predictions for these probabilities, which we can compare with the measurements.
\begin{figure}[!h]
\includegraphics[scale=0.4]{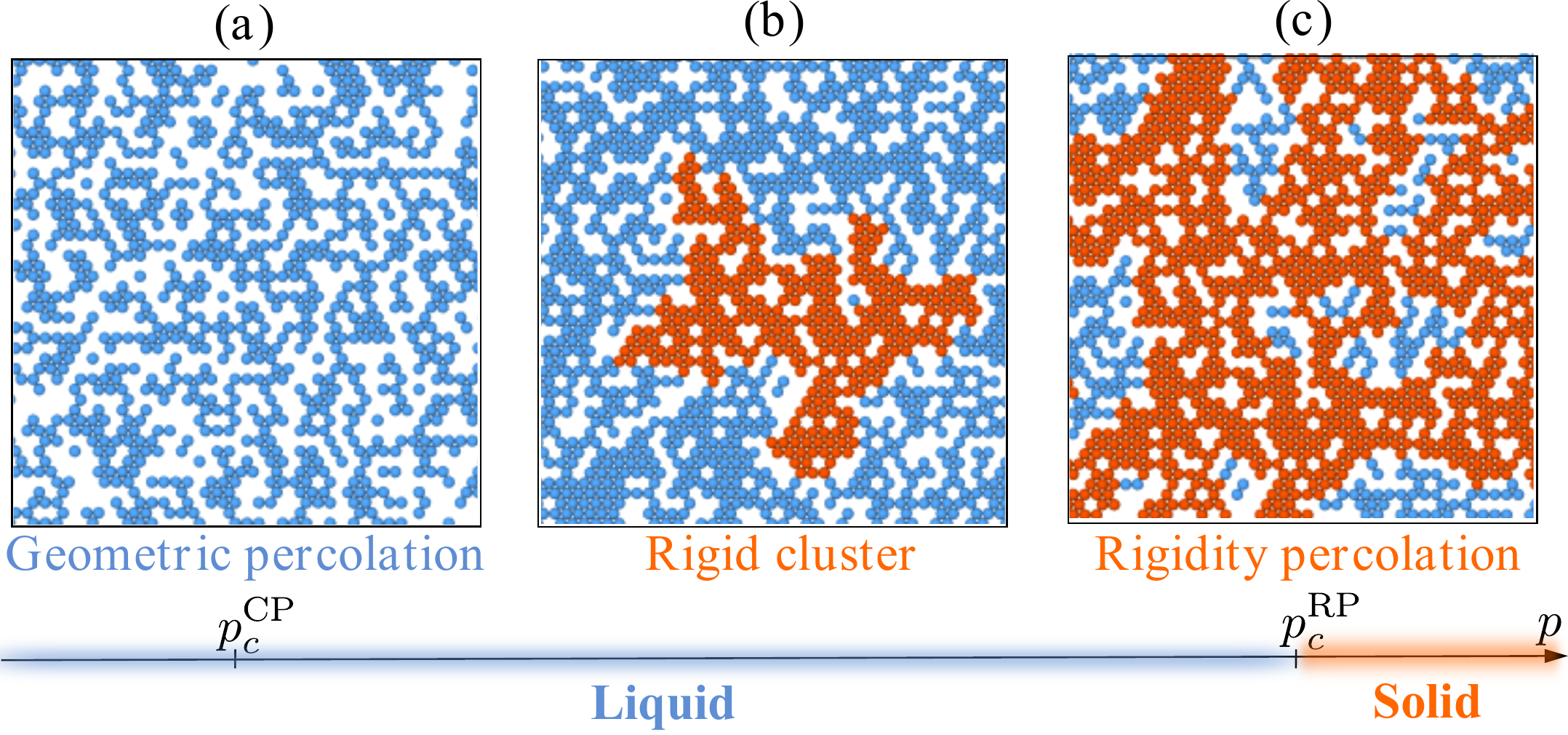}
\caption{Examples of site-diluted triangular lattice configurations showing connectivity percolation transition (a) at $p_c^{CP}$ for which the system is macroscopically liquid. (b) Rigid cluster decomposition where red particles belong to the largest rigid cluster obtained via constraints counting analysis (pebble game) and (c) macroscopic rigidity percolation transition at $p_c^{RP}$  exhibiting a percolating rigid cluster in the two directions able to sustain external loads.}
\label{fig:epsart} 
\end{figure}
The model is a site-diluted triangular lattice with local spatial correlations. It has been recently introduced to model the rigidity percolation of soft solids~\cite{mehdi19}. At each step, particles are drawn randomly one by one to populate a doubly-periodic triangular lattice of size $L_1\times L_2$, according to the following
probability $p=(1-c)^{6- N_n}$, where $c\in[0,1[$ represents the degree of correlation and $N_n $ is the number of nearest filled sites varying between $0$ to $6$ for fully occupied neighboring sites. Since the filling probability depends only on the degree of occupation of the first neighbors, the introduced correlations are local and in the limit of $c=0$ we recover the classical uncorrelated random percolation where all particles has the same filling probability. In practice, the larger $c$ the smallest is the critical probability threshold $p_c^{RP}$ (equivalently critical volume fraction) which yields to macroscopic finite elasticity. These correlations are irrelevant and the large-scale behaviour is unaffected by the value of $c$, so that the transition still belongs to the same universality class as classical uncorrelated RP~\cite{mehdi19}. In practice we used $c=0.3$ at which $p_c^{RP}(c=0.3) \sim 0.66$. 

To identify rigid clusters on a discrete lattice, we use the so-called 'Pebble game', a fast combinatory algorithm\cite{jacobs1995generic,jacobs96}. It is based on Laman's theorem for graphs' rigidity, which uses Maxwell's constraint counting argument for each subgraph to detect over-constrained clusters highlighting rigidity \cite{laman1970graphs}. Figure~\ref{fig:epsart} shows an example of cluster decomposition while increasing $p$. Connectivity percolation arises at $p_c^{\rm CP}$ and is characterized by a space-spanning percolating cluster (in blue). The system is macroscopically liquid and cannot sustain external loads. Figure~\ref{fig:epsart}b and \ref{fig:epsart}c show the largest rigid cluster (in red) that percolates at $p_c^{\rm RP}>p_c^{\rm CP}$, leading to macroscopic elasticity.
\\
In the following, we analyse the statistical properties of the rigid clusters at the critical point. We first obtain a direct measurement of the anomalous dimension exponent $\eta$, then move on to test conformal invariance, using the 3-point and 2-point connectivities. Finally we highlight the similarities with CP in the structure of these functions.

\begin{figure}[htb]
\includegraphics[]{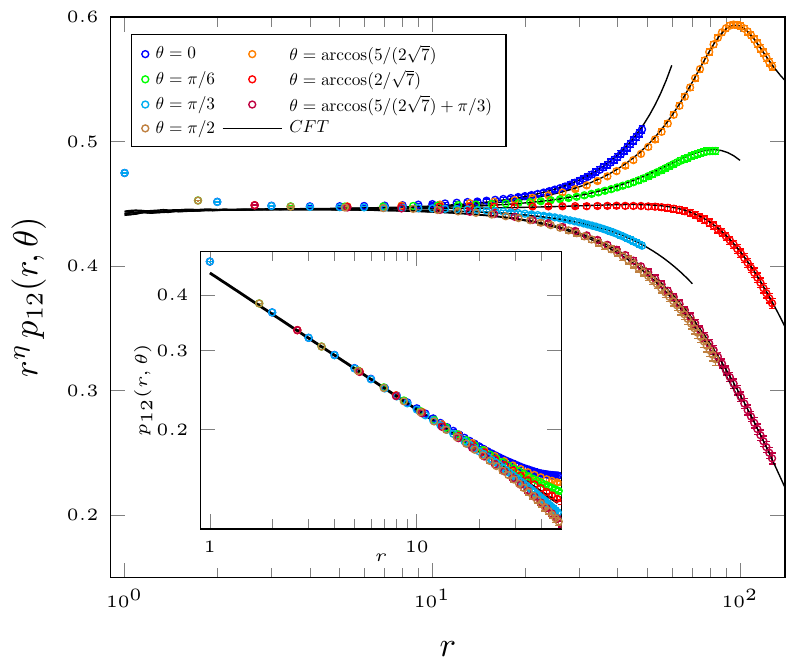}
\caption{Rescaled 2-point connectivity measured in the cylinder limit $L_1=6L_2$, $L_2 = 96$ along different angles, in semilog scale. The black curves are the CFT prediction (\ref{eq:2ptcyl}). Inset: the same data points, not rescaled, in log-log scale. The black line has slope $\eta = 0.307$.}
\label{fig:local}
\end{figure}

\textit{Anomalous dimension --}
We measure the 2-point connectivity $p_{12}(r,\theta)$ on the lattice, ie the probability that points $(i,j)$ and $(i+r \cos(\theta+\pi/3),j+r \sin(\theta+\pi/3))$ are in the same rigid cluster. $\theta$ is the angle wrt the short cycle of the doubly-periodic lattice, and $r$ the distance between the two points. We use translation invariance to average over the $L_1\times L_2$ positions $(i,j)$, as well as symmetry by reflection about $\theta=0$, so that $p_{12}$ is an average over $2 L_1 L_2 N$ measurements with $N$ the number of samples ($N=1200$ for the largest sizes). The inset in figure \ref{fig:local} shows the data points in log-log scale which follow a power law in the scaling region $1\ll r \ll L_2/2$. This is expected from scale invariance, namely that for points separation $1\ll z_{12} \ll L_2$, the 2-point connectivity decays as $p_{12}(z_1,z_2)\sim \left|z_{12}\right|^{-\eta}$, where $\eta$ is the so-called anomalous dimension, satisfying the hyperscaling relations $\eta = 2\beta/\nu = 4-2d_f$ \cite{SA92}. Using assumption (\ref{eq:assumption}) and that $\langle \Phi_c(z_1)\Phi_c(z_2)\rangle =z_{12}^{-2\Delta_c}$ \cite{DiFrancesco:1997nk}, gives the scaling dimension of $\Phi_c$ as $\Delta_c = \eta/2$.
 Expected deviations in the region $r\sim L_2/2$ are due to universal finite size effects coming from the doubly-periodic boundary conditions. Fitting the data points corresponding to the angle that minimises such effects ($\theta= \arccos(2/\sqrt(7))$) we obtain the value of the non-universal constant $a_0^{(2)} = 0.448\pm0.002$, and of the anomalous dimension $\eta = 0.307\pm0.002$, in agreement with the values of the critical exponents in the literature \cite{jacobs1995generic} via hyperscaling.

\begin{figure}[htb]
 \includegraphics[]{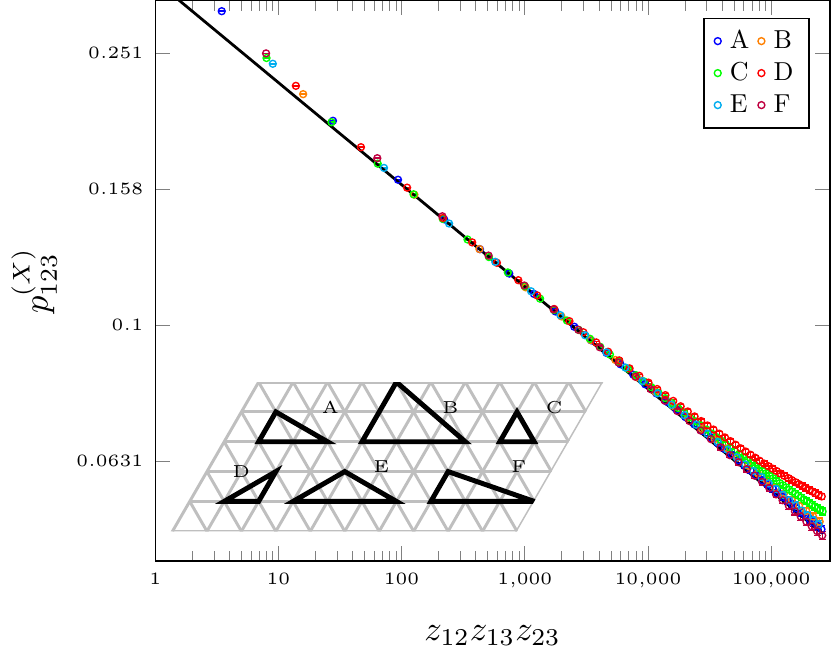}
\caption{ 3-point connectivity measured on the six inequivalent triangles shown in the inset, on a $L_2\times 3 L_2$ lattice with $L_2=2^7$.}
\label{fig:global}
\end{figure}

\textit{Global conformal invariance --}
In two dimensions, conformal transformations are all the analytic maps on the Riemann sphere (complex plane plus point at infinity). They can be distinguished into a finite set of globally defined (everywhere invertible) transformations (translation, rotation, scaling, special conformal transformation), and an infinite set of local transformations (see eg. \cite{DiFrancesco:1997nk}).
It is a standard result that imposing invariance under the global tranformations fixes completely the form of 3-point correlations, so that, using (\ref{eq:assumption}) one expects the 3-point connectivity of globally invariant clusters to be \cite{DiFrancesco:1997nk}:
\begin{equation}\label{eq:3ptconf}
p_{123}(z_1,z_2,z_3) = a_0^{(3)}\frac{C_{\Phi_c\Phi_c}^{ \Phi_c}}{\left|z_{12}z_{23}z_{13}\right|^{\eta/2}}
\end{equation}
where $C_{\Phi_c\Phi_c}^{\Phi_c}$ is an universal constant called operator product expansion (OPE) coefficients (see eg \cite{Cardybook}). Note that for a scale but not conformal invariant system we expect instead
\begin{equation}\label{eq:3ptSI}
\begin{aligned}
p_{123}(z_1,z_2,z_3) = &\frac{a_0^{(3)}}{\left|z_{12}z_{23}z_{13}\right|^{\eta/2} }\sum_{a+b+c = 0}\frac{C^{(abc)}_{\Phi_c \Phi_c\Phi_c}}{\left|z_{12}\right|^a\left|z_{23}\right|^b\left|z_{13}\right|^c}\\
&+ \mathrm{perm.}\left[1\leftrightarrow2,\,1\leftrightarrow3,\,2\leftrightarrow3\right].
\end{aligned}
\end{equation}
In Figure \ref{fig:global} we show $p_{123}$ measured on 6 inequivalent configurations of points, plotted as a function of $z_{12}z_{13}z_{23}$. A clear collapse is seen in the scaling region $1\ll z_{ij} \ll L_2/2$, showing the validity of (\ref{eq:3ptconf}), while (\ref{eq:3ptSI}) cannot hold. Microscopic and configuration-dependent finite-size effects dominate at small and large separation respectively.

\textit{Local conformal invariance --}
We now use the universal finite-size effects induced by the torus geometry (doubly-periodic bc) to probe the local conformal invariance. In particular, we put the system on a cylinder, conformally equivalent to the plane through the map $z\to i L_2/2\pi\,\log z$. For a CFT, the expression of a 2-point correlation function on this geometry is a well-known result \cite{DiFrancesco:1997nk}, which in terms of the 2-point connectivity and in polar coordinates reads:

\begin{equation}\label{eq:2ptcyl}
     p_{12}(r,\theta) =  \frac{a_0^{(2)}\left(2\pi/L_2\right)^\eta}{\left[2\cosh(\frac{2\pi}{L_2}r\cos\theta)-2\cos(\frac{2\pi}{L_2}r\sin\theta)\right]^{\eta/2}}
\end{equation}
This prediction is drawn in figure \ref{fig:local} for different angles $\theta$, using the values of $\eta$ and $a_0^{(2)}$ found previously, along with the corresponding numerical data points measured on a torus with large aspect ratio $L_1/L_2 = 6$ to reproduce the cylinder limit. The remarkable agreement confirms that the 2-point connectivity of rigid clusters transforms correctly under this local conformal transformation. In more technical terms, the data is consistent with the connectivity field $\Phi_c$ being a Virasoro primary, so that one can expect all connectivities to be conformally invariant as well.

\begin{figure}[!h]
     \includegraphics[]{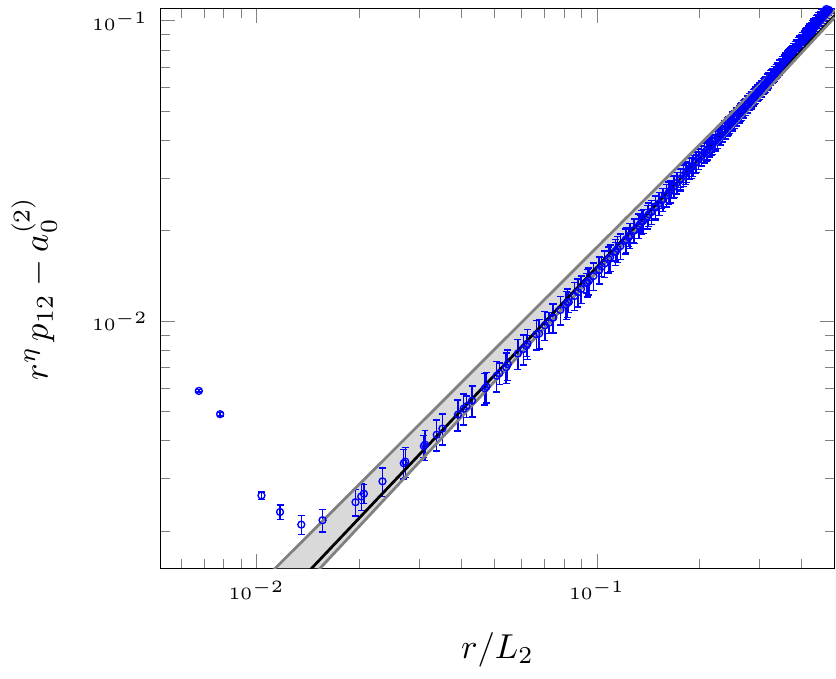}
\caption{Blue points correspond to measurements on a lattice of size $L_2 = 256$. The black line gives the best fit, done in the range $1\ll r\ll L_2$, and yields $\nu=1.19\pm0.01$. The gray area corresponds to $\nu$ in the confidence interval of ref.\cite{jacobs1995generic} , $\nu\in[1.15,1.27]$.}
\label{fig:thermal}
\end{figure}

\begin{figure}
\centering
\includegraphics[]{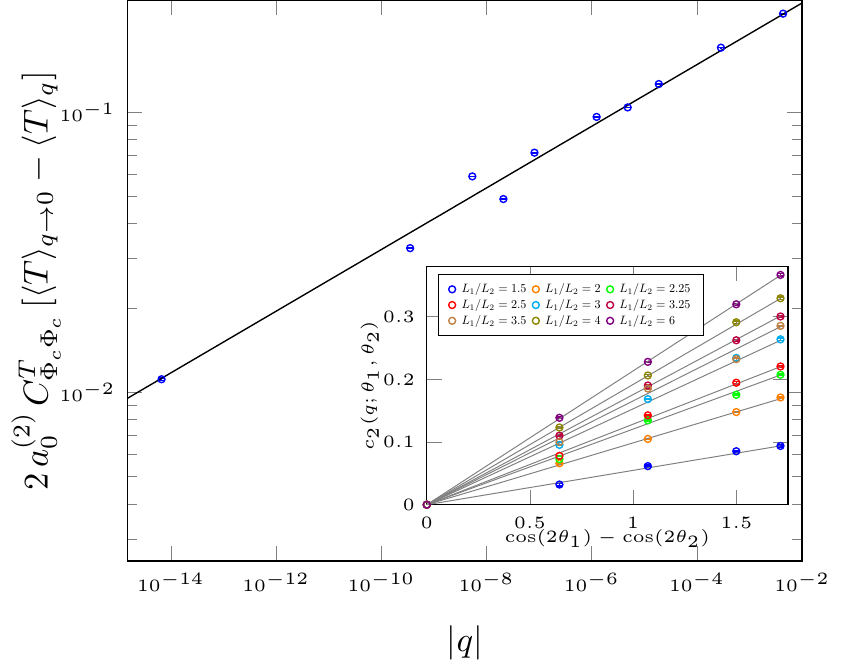}
\caption{
Behaviour of the 1-point function of the stress-energy tensor with $q$. The black line corresponds to the fit $\sim\left|q\right|^{0.11}$.
Inset: order 2 coefficients $c_2$ for different aspect ratio. The grey lines are the best fits, whose slopes give the points of the main plot.}
\label{fig:T}
\end{figure}

\textit{Finite-size corrections and comparison with CP}
On a doubly-periodic system of finite aspect ratio, one can write generically a so-called OPE expansion, for $r\ll L_2$, of the 2-point connectivity as \cite{Javerzat2019}:
\begin{equation}\label{eq:p12toregen}
\begin{aligned}
    p_{12}(r,\theta)& = \frac{a_0^{(2)}}{r^\eta}\Bigg(1+\\
    &\sum_{\Phi_\alpha}C_{\Phi_c\Phi_c}^{\Phi_\alpha}\langle \Phi_\alpha\rangle_q\, (2-\delta_{s_\alpha,0})\cos(s_\alpha\theta)\left(\frac{r}{L_2}\right)^{\Delta_\alpha}\Bigg)
    \end{aligned}
\end{equation}
The sum is a (potentially infinite) sum over an a priori unknown set of fields $\Phi_\alpha$ with dimension $\Delta_\alpha$ and spin $s_\alpha\geq0$. Each field contribution gives a $r/L_2$ correction of order $\Delta_\alpha$ to the plane limit, and depends on the elliptic nome of the torus $q=e^{-2\pi L_1/L_2\sin\pi/3}e^{2\pi i L_1/L_2\cos\pi/3}$ through the expectation value of $\phi$ on the torus, and on $\theta$ for non-scalar fields (of non-zero spin). Note that on a square torus ($L_1=L_2$) $p_{12}$ is independent of $\theta$ and the expectation of non-scalar fields must vanish.
We find that, for RP, the first terms in expansion (\ref{eq:p12toregen}) are the following:
\begin{equation}\label{eq:p12toreC}
\begin{aligned}
    \frac{r^\eta}{a_0^{(2)}} p_{12}(r,\theta) &= 1+C_{\Phi_c\Phi_c}^{\Phi_\nu}\langle \Phi_\nu\rangle_q\left(\frac{r}{L_2}\right)^{2-1/\nu}\\
    &+2C_{\Phi_c\Phi_c}^{T}\langle T\rangle_q\cos(2\theta)\left(\frac{r}{L_2}\right)^2+\cdots
    \end{aligned}
\end{equation}
Namely, the dominant finite-size correction is given by the (scalar) "thermal" field $\Phi_\nu$ whose dimension is $\Delta_\nu = 2-1/\nu$, and the first non-scalar contribution comes from the so-called stress-energy tensor $T$. This latter field is the tensor of conserved currents arising from translation invariance, and from dimensional analysis have dimension 2 and spin 2. The dots account for the higher order, unknown contributions. \\
In figure \ref{fig:thermal} we show the dominant finite-size correction: by measuring $p_{12}$ on a square torus we eliminate the non-scalar contributions to (\ref{eq:p12toreC}), so that the quantity $r^\eta p_{12}-a_0^{(2)}$ is directly proportional to the dominant scalar contribution, up to subleading corrections. The grey area corresponds to a term $\sim\left(r/L_2\right)^{2-1/\nu}$ with $\nu$ in the confidence interval of ref \cite{jacobs1995generic}, $\nu = 1.21\pm0.06$, showing that the data is consistent with $\Delta_{\mathrm{dominant}} = 2-1/\nu = \Delta_\nu$. Fitting in the range $1\ll r \ll L_2/2$ gives $\nu = 1.19\pm0.01$.\\
The dominant non-scalar field contribution is instead obtained by getting rid of the scalar terms in (\ref{eq:p12toreC}), which is achieved by measuring $p_{12}$ in two directions $\theta_1,\,\theta_2$, and is consistent with an order 2 term, namely:
\begin{equation}
\begin{aligned}\label{eq:order2}
    r^\eta \,\Big[ p_{12}&(r,\theta_1)- p_{12}(r,\theta_2)\Big] = \\
    &\underbrace{a_0^{(2)}2C_{\Phi_c\Phi_c}^{T}\,\langle T\rangle_q \left[ \cos(2\theta_1)-\cos(2\theta_2)\right]}_{\equiv c_2(q;\theta_1,\theta_2)}\left(\frac{r}{L_2}\right)^2+\cdots
    \end{aligned}
\end{equation}
We extracted the order 2 coefficients $c_2(q;\theta_1,\theta_2)$ of $r^\eta\left[ p_{12}(r,\theta_1)- p_{12}(r,\theta_2)\right]$, measured for different aspect ratios and different angles, and plotted them in the inset of figure \ref{fig:T} as a function of $\cos(2\theta_1)-\cos(2\theta_2)$. The nice straight lines confirm that we are indeed measuring the contribution of a dimension 2 and spin 2 field, ie of $T$. From (\ref{eq:order2}) their slopes correspond to $a_0^{(2)}2C_{\Phi_c\Phi_c}^{T}\,\langle T\rangle_q$, and are plotted as function of the elliptic nome $q$ in figure \ref{fig:T}. Fitting these points we find that 
\begin{equation}\label{eq:fitT}
    \langle T\rangle_{q\to0}-\langle T\rangle_q\sim \left|q\right|^{\Delta_{0}},\quad \Delta_{0}\sim0.11.
\end{equation}
First, this form is consistent with CFT, which gives $\langle T\rangle_q$ as \cite{DiFrancesco:1997nk}
\begin{equation}\label{eq:oneptT1}
    \langle T \rangle_q = -(2\pi)^2 q\partial_q \log Z(q) .
\end{equation}
$Z(q)$ is the so-called partition function on the torus, $Z(q)\equiv \sum_{\Phi_\alpha} n_\alpha q^{(\Delta_\alpha+s_\alpha)/2-c/24}\bar{q}^{(\Delta_\alpha-s_\alpha)/2-c/24}$, with $c$ the so-called central charge --an important parameter characterising a CFT, and $n_{\alpha}$ the multiplicity of field $\Phi_\alpha$. Expanding (\ref{eq:oneptT1}) for small $q$ gives
\begin{equation}\label{eq:oneptT2}
\begin{aligned}
    \langle T \rangle_q\overset{q\ll 1}{=}&(2\pi)^2\Big[\frac{c}{24}\\
    &-\sum_{\Phi_\alpha} n_\alpha \frac{\Delta_\alpha+s_\alpha}{2} q^{\frac{ \Delta_\alpha+s_\alpha}{2}}\bar{q}^{\frac{\Delta_\alpha-s_\alpha}{2}}+\cdots\Big]
    \end{aligned}
\end{equation}
The constant term corresponds to the cylinder limit $\langle T \rangle_{q\to0} = (2\pi)^2c/24$. Equation (\ref{eq:oneptT1}) is actually one of the most direct consequences, at the level of observables, of conformal invariance, coming from the holomorphicity of $T$ ie $\bar{\partial}T = 0$.\\
Secondly, from (\ref{eq:fitT}) the value of the smallest dimension in the sum (\ref{eq:oneptT2}), denoted $\Delta_{0}$, is compatible with the scaling dimension of $\Phi_c$, $\Delta_{c} = \eta/2 = 0.15\pm0.02$, and so compatible with the connectivity field $\Phi_c$ being the field with smallest non-zero scaling dimension in the theory. Namely
\begin{equation}\label{eq:oneptT3}
    \langle T_q\rangle = (2\pi)^2\left[\frac{c}{24}-n_{c} \eta \left|q\right|^{\eta/2}+\cdots\right].
\end{equation}
\\
It has been established that the expansion  (\ref{eq:p12toreC}) of the torus 2-point connectivity is valid for CP models, in \cite{Javerzat2019} for the $Q-$state Potts model and in \cite{javerzat2020topological} for percolation of random surfaces, two families of correlated percolation models which include uncorrelated CP as a limiting case and span a continuum of universality classes. Namely it was found that the dominant terms in the $\Phi_c\times\Phi_c$ OPE are the conformal families of the identity and thermal fields. For these models it was also found that the field with smallest scaling dimension entering the partition function is the connectivity field.
Therefore, our results indicate that --within our numerical range, the structures of the 2-point connectivity and of the torus partition function are identical in RP and in CP. In other words, at the level of the geometry of random clusters there is not more difference between RP and CP than between two different CP universality classes. In this respect it would be useful to characterise more precisely the CFT of RP clusters, by determining in particular its central charge $c$. This data is not accessible in our study as it cancels in (\ref{eq:order2}), given that $C_{\Phi_c\Phi_c}^T =\Delta_c/c$ (see eg. \cite{ninathese}).

\textit{Conclusion -- }
In this work, we have investigated the rigidity percolation transition. Through a series of three original and independent tests, we have shown -- for the first time -- that the statistical properties of the random fractal clusters, encoded in the connectivity functions, are conformally invariant at the critical point. Given that RP exhibits highly non-local interactions --where the removal of a single bond might destroy the rigidity of an arbitrary large region-- it is quite remarkable that invariance under local rescalings holds, and that one can predict the cluster connectivity properties using correlations of local fields. \\ 
Surprisingly, we found that the structure of the connectivity functions is identical to what we expect for connectivity percolation, albeit with a priori different values of the universal data (critical exponents and OPE coefficients). Therefore, although it is widely believed that the rigidity and the connectivity percolation phenomena are of fundamentally different natures, our work provides evidences on the similarity of their clusters at criticality. These findings support the suggestion of \cite{Head2003}, that the geometrical properties of rigidity might be physically independent from the elastic properties. Recent work on the RP for granular media near jamming transition~\cite{liu2019frictional} also points towards a possible superuniversality of some RP and CP critical exponents.
Many questions remain thus open, it would be interesting to extend our approach to probe the signature of conformal invariance in the mechanical behaviour of RP, by studying eg. the stress transmission at the verge of rigidity. Indeed while recent field theories have been very successful in predicting the elastic response in disordered amorphous materials~\cite{degiuli2018field,nampoothiri2020emergent} away from the critical point, the vicinity of the transition is much less understood. In this respect, our approach of using conformal field theory may open a new avenue of thinking to build a unified framework to describe the mechanical properties of a wide range of materials close to their rigidity transition.

\begin{acknowledgments}
The authors thank Alexandre Nicolas, Filiberto Ares, Xiaoming Mao, Emanuela Del Gado and Ezequiel Ferrero for insightful discussions, as well as Raoul Santachiara who suggested this problem to us.
\end{acknowledgments}

\bibliography{apssamp}

\end{document}